\documentclass[cits]{PoS}
\usepackage{amssymb,amsfonts,amsmath,bm,bbm,fontenc,times,mathptmx,graphicx}

\newcommand{\mf}[1]{\mathbf{#1}}
\newcommand{\ov}[1]{\overline{#1}}
\newcommand{\be}{\begin{equation}}
\newcommand{\ee}{\end{equation}}
\newcommand{\bey}{\begin{eqnarray}}
\newcommand{\eey}{\end{eqnarray}}

\newcommand{\E}{\mathrm{e}}

\title{The chiral and angular momentum content of the $\bm\rho$-mesons in
lattice QCD}

\ShortTitle{The chiral and angular momentum content of the $\rho$-mesons in
lattice QCD}

\author{L.~Ya.~Glozman\\
        Institut f\"ur Physik, FB Theoretische Physik,
        Universit\"at Graz, A--8010 Graz, Austria\\
        E-mail: \email{leonid.glozman@uni-graz.at}}
\author{C.~B.~Lang\\
        Institut f\"ur Physik, FB Theoretische Physik,
        Universit\"at Graz, A--8010 Graz, Austria\\
        E-mail: \email{christian.lang@uni-graz.at}}
\author{\speaker{Markus Limmer}\\
        Institut f\"ur Physik, FB Theoretische Physik,
        Universit\"at Graz, A--8010 Graz, Austria\\
        E-mail: \email{markus.limmer@uni-graz.at}}

\abstract{The variational method allows one to study the mixing of interpolators
with different chiral transformation properties in the nonperturbatively
determined physical state. It is then possible to define and calculate in a
gauge-invariant manner the chiral as well as the partial wave content of the
quark-antiquark component of a meson in the infrared, where mass is generated.
Using a unitary transformation from the chiral basis to the $^{2S+1}L_J$ basis
one may extract the partial wave content of a meson. We present results for the
$\rho$- and $\rho'$-mesons using a simulation with $N_f=2$ dynamical quarks, all
for lattice spacings close to $0.15$ fm. Our results indicate a strong chiral
symmetry breaking in the $\rho$ state and its simple $^3S_1$-wave composition 
in the infrared. For the $\rho'$-meson we find a small chiral symmetry breaking
in the  infrared as well as a leading contribution of the $^3D_1$ partial wave,
which is contradictory to the quark model.}

\FullConference{The XXVIII International Symposium on Lattice Field Theory,
Lattice2010\\ June 14-19, 2010\\ Villasimius, Italy}

\begin{document}

\section{Introduction}

\noindent
A central question in QCD is the mass generation mechanism and its
interconnection with confinement and chiral symmetry breaking.  The angular
momentum generation of hadrons is another related question. Chiral symmetry is
dynamically broken in the QCD vacuum which is evidenced by the absence of 
parity doublets in the low-lying hadron spectrum and by the existence of the
pion as a pseudo Goldstone boson. It follows from the trace anomaly that the
hadron mass almost completely consists of the energy of the quantized gluonic
field. However, this  statement tells us nothing about mechanism of mass
generation. It is widely believed that the spontaneous breaking of chiral
symmetry, i.e., the quark condensate of the vacuum, is prominently  responsible
for the mass generation of hadrons like the nucleon or $\rho$-meson. This was
observed in various microscopical models and in  the QCD sum rule approach
\cite{SVZ,I}.

Chiral symmetry breaking in the vacuum  explains a phenomenological success of
the quark model, at least for the ground states. Namely, almost massless light
quarks acquire their effective (dynamical, constituent) mass at low momenta via
their coupling with the quark condensate. This large mass renders the problem
effectively non-relativistic and the ground state $\rho$ is a $^3S_1$ state in
the quark model language \cite{PDG,IS}. Traditionally the excitation spectrum is
also described within the quark model and the first excited state of the
$\rho$-meson, $\rho(1450)$, is believed to be the first radial excitation,
i.e.,  a $^3S_1$ state \cite{PDG,IS}.

At the same time there are indications that chiral symmetry is effectively
restored in the high-lying spectrum \cite{Gloz1,Gloz2}. This would imply that
the mass generation mechanism in highly  excited hadrons is different and the
quark condensate of the vacuum is of little importance. It would also imply that
the constituent quark model language is inadequate for excited hadrons. To
resolve the issue one needs direct information about the hadron structure, which
can be supplied in ab initio lattice simulations. Here we present a way to
reconstruct in dynamical simulations a chiral as well as an angular momentum
decomposition of the leading quark-antiquark component of mesons at physical,
infrared scale.

The variational method \cite{Mi85} provides a tool to study the hadron wave
function in lattice QCD calculations.  One uses a set of interpolators
$\{O_1,O_2,\ldots,O_N\}$, which have the quantum numbers of the state of
interest, and computes the cross-correlation matrix,
\be\label{c_ij}
C_{ij}(t) = \big\langle\  O_i(t)\  O_j^\dag (0)\ \big\rangle\ .
\ee
One solves a generalized eigenvalue problem;  assuming that the set of
interpolators $\{O_i\}$ is complete enough,  the wave function is related to the
eigenvectors obtained. We are interested in the reconstruction of the leading
quark-antiquark component of the low lying mesons. Therefore we need
interpolators that allow us to define such a component in a unique way.

In \cite{Gloz1} a classification of all non-exotic quark-antiquark states
(interpolators) in the light meson sector according to the transformation
properties with respect to the $SU(2)_L\times SU(2)_R$ and $U(1)_A$ was
presented.  If no explicit excitation of the gluonic field with non-vacuum
quantum numbers is present, this basis is a complete one for a quark-antiquark
system and we can define and investigate chiral symmetry breaking. Namely, one 
can reconstruct a decomposition for a given meson  in terms of different
representations of the chiral group by diagonalizing the cross-correlation
matrix from \eqref{c_ij}. The eigenvectors describe the quark-antiquark content
in terms of different chiral representations. If we observe components with
different transformation properties in terms of $SU(2)_L\times SU(2)_R$ and
$U(1)_A$, then we conclude that chiral symmetry is broken in that state.

In order to establish a connection to the quark model, it is interesting to
reconstruct the meson composition in terms of the ${}^{2S+1}L_J$ basis, where
$\mf J = \mf L + \mf S$ are the standard angular momenta in the two-body
system.  Such a  decomposition of the leading quark-antiquark component in terms
of the ${}^{2S+1}L_J$ basis in the infrared, i.e., where the hadron mass is
generated, would tell us to which degree the quark model language is adequate
for a given state.

The ${}^{2S+1}L_J$ angular momentum basis and the chiral basis are both complete
for a two-particle system. They are connected  to each other via  a unitary
transformation \cite{GN}. Each state of the chiral basis can be uniquely
represented in terms of the ${}^{2S+1}L_J$ states. Then, diagonalizing the
cross-correlation matrix, built from interpolators with definite chiral
transformation properties, one can obtain the partial wave decomposition of the
leading Fock component.  This method can in principle be applied to any meson,
here we use as an example the vector meson $\rho$ and its first excitation
$\rho(1450)$  \cite{GLL,GLL2}.

\section{Chiral classification and the transformation to the angular momentum basis}

\noindent
The classification of the quark-antiquark states and  interpolators with respect
to representations of $SU(2)_L \times SU(2)_R$ was done in \cite{Gloz1}.  We are
interested in the quark-antiquark component of the ground state $\rho$-meson and
its first excitation. There are two possible chiral representations that are
compatible with the  quantum numbers of  the $\rho$-meson, which have
drastically different chiral transformation properties. Assuming that chiral
symmetry is not broken, then one has two independent states. The first state is
$\vert(0,1)\oplus (1,0);\, 1\, 1^{--}\rangle$; it can be created from the vacuum
by the standard vector current 
\be\label{eq:o_v}
O_V = \ov{q} \,\gamma^i\, \vec{\tau}\, q\  .
\ee
Its chiral partner is the $a_1$ meson. The other state is $\vert(1/2,1/2)_b;\,
1\, 1^{--}\rangle$, which can be created by the pseudotensor operator,
\be\label{eq:o_t}
O_T = \ov{q} \,\sigma^{0i} \,\vec{\tau}\, q\ ,
\ee
and its chiral partner is the $h_1$ meson. Here, $\vec{\tau}$ denotes the vector
of isospin Pauli matrices.

Chiral symmetry breaking in the state implies that the state should be a mixture
of both representations. If it is a superposition of both representations with
approximately equal weights, then the chiral symmetry is maximally violated in
this state. If, on the contrary, one of the representations strongly dominates
over the other representation, one could speak about effective chiral 
restoration in this state.

These chiral representations can be transferred into the  ${}^{2S+1}L_J$ basis,
using the unitary transformation \cite{GN}
\be\label{eq:trafo}
\left( \begin{array}{c}
\vert(0,1)\oplus (1,0);\, 1\, 1^{--}\rangle\\
\vert(1/2,1/2)_b;\, 1\, 1^{--}\rangle\\
\end{array} \right)
=
U\cdot
\left( \begin{array}{c}
\vert 1;\, {}^3S_1\rangle\\
\vert 1;\, {}^3D_1\rangle\\
\end{array} \right)\ ,
\ee
where $U$ is given by
\be
U=\left( \begin{array}{lr}
\sqrt{\frac{2}{3}}\ &\  \sqrt{\frac{1}{3}}\\
\sqrt{\frac{1}{3}}\ &\ -\sqrt{\frac{2}{3}}\\
\end{array} \right)\ .
\ee

Thus, using the interpolators $O_V$ and $O_T$ from \eqref{eq:o_v} and
\eqref{eq:o_t} for the diagonalization of the cross-correlation matrix, we are
able to reconstruct the partial wave content of the leading  $\bar q q$-Fock
component of the $\rho$-meson.

\section{Reconstruction of the wave function using the variational method}

\noindent We briefly want to discuss the basic features of the variational
method \cite{Mi85} and how to analyse the decomposition of the $\rho$-mesons.
The time propagation properties of the normalized physical states $\vert
n\rangle$ are given by
\be
\langle n(t) \vert m(0) \rangle = \delta_{nm}\, \E^{-E_n \,t}\ .
\ee
The lattice interpolators $O_i$ are typically not normalized and are projected
to zero spatial momentum. The cross-correlation matrix from Eq.\ \eqref{c_ij}
can be written as
\be
C_{ij}(t) = \big\langle\, O_i(t)\,O_j^\dag (0)\,\big\rangle\
          = \sum_n\, a_i^{(n)} \,a_j^{(n)*}\, \E^{-E_n \,t}\ ,
\ee
where the coefficients $a_i^{(n)}$ give us the overlap of the physical state
$\vert n\rangle$ with the lattice interpolator $O_i$,
\be
a_i^{(n)} = \langle 0 \vert O_i \vert n \rangle\ .
\ee
The two chiral representations $(0,1)\oplus (1,0)$ and $(1/2,1/2)_b$ form a
complete and orthogonal basis (with respect to the chiral group) for
$\rho$-mesons. Consequently, using the variational method we are able to study
the mixing of the two representations in both $\rho$ and $\rho'$ states.

Following the lines of \cite{GLL} one can show that the ratio of couplings can
be obtained as
\be\label{eq:ratio}
\frac{a_i^{(n)}(t)}{a_k^{(n)}(t)} =
\frac{C_{ij}(t)\, u_j^{(n)}(t)}{C_{kj}(t)\, u_j^{(n)}(t)}\ .
\ee
The ratio of the vector to pseudotensor couplings, $a_V^{(n)}/a_T^{(n)}$, tells
us about the chiral symmetry-breaking in the states $n=\rho,\rho'$.

\section{Defining the resolution scale}

\noindent
If we probe the hadron structure with the local interpolators, then we study the
hadron decomposition at the scale fixed by the lattice spacing $a$. For a
reasonably small $a$ this scale is close to the ultraviolet scale. However, we
are interested  in the hadron content at the infrared scales, where mass is
generated. For this purpose we cannot use a large $a$, because matching with the
continuum QCD will be lost. Given a fixed, reasonably small lattice spacing $a$ a
large infrared scale $R$ (i.e., small resolution scale $1/R$) can be achieved by
gauge-invariant smearing of the point-like interpolators. We smear the quark
field (sources) in spatial directions  over the size $R$ in physical units, such
that $R/a\gg 1$. Then, even in the continuum limit $a \rightarrow 0$ we probe the
hadron content at the infrared scale fixed by $R$. Such a definition of the
resolution is similar to the experimental one, where an external probe is
sensitive only to quark fields (it is blind to gluonic fields) at a resolution 
that is determined by the momentum transfer in spatial directions.

The smearing procedure itself is done using Jacobi smearing \cite{GuskenBest}. A
smearing operator $M$ acts on a point-like source $S_0$,
\be
S = M\,S_0\ ,\qquad M=\sum_{n=0}^N (\kappa \,H)^n\ .
\ee
The hopping term $H$ is given by
\be
H=\sum_{k=1}^3 \big[ U_k(x,t)\,\delta_{x+\hat{k},y} +
U_k^\dag(x-\hat{k},t)\,\delta_{x-\hat{k},y} \big]\ .
\ee
It creates approximately a Gaussian profile of the width $R$ for each quark
field of the smeared interpolator.

\section{Lattice simulation details and results}

\begin{table}[tb]
\begin{center}
\begin{tabular}{ccccccc}
\hline
\hline
Set&  $\beta_{LW}$ &  $a\,m_0$ & \#{conf} & $a$ [fm] & $m_\pi$ [MeV] &
$m_\rho$ [MeV]\\
\hline
A & 4.70 & -0.050 & 200 & 0.1507(17) & 526(7) & 911(11) \\
B & 4.65 & -0.060 & 300 & 0.1500(12) & 469(5) & 870(10) \\
C & 4.58 & -0.077 & 300 & 0.1440(12) & 323(5) & 795(15) \\
\hline
\hline
\end{tabular}
\caption{Details of the lattice simulation: the leading value $\beta_{LW}$ of
the gauge coupling, the bare mass parameter $a\,m_0$ of the CI action, the
number of analyzed configurations, the lattice spacing $a$, the pseudoscalar
mass $m_\pi$, and the vector meson mass $m_\rho$ (cf.\ \cite{hadron2} for more
details).}
\label{tab_data}
\end{center}
\end{table}

\noindent
Like in our previous  analysis of excited hadrons \cite{hadron1,hadron2}, we use
the L\"uscher-Weisz gauge action \cite{LuWe85} and the Chirally Improved (CI)
Dirac operator, which has better chiral properties than the Wilson Dirac
operator \cite{Ga01a}. For this study three sets of dynamical gauge
configurations, all for lattice size of $16^3\times 32$, including two
mass-degenerate light sea quarks are used (for details see Tab.\
\ref{tab_data}).

\begin{figure}[tb]
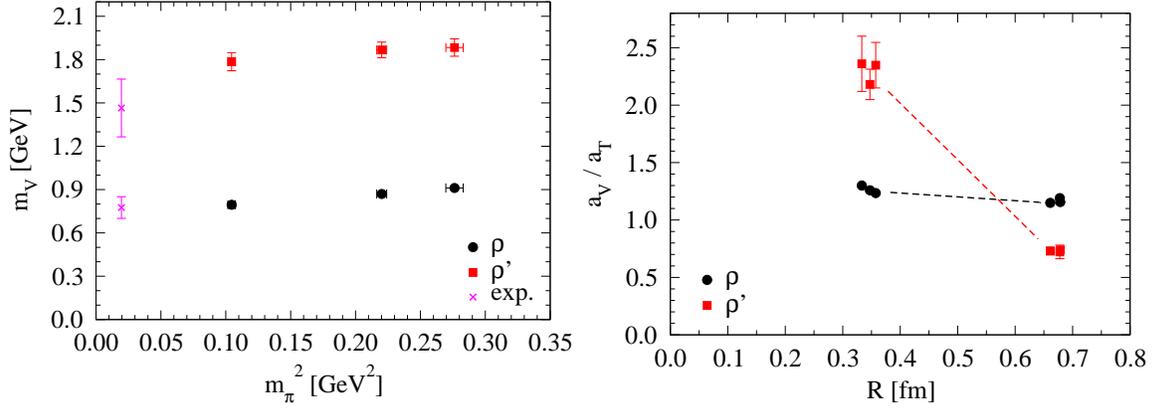

\begin{center}
\includegraphics[width=75mm,clip]{./mass_rho.eps}\hfill
\includegraphics[width=75mm,clip]{./ratio_vs_R.eps}
\caption{L.h.s.: The vector meson mass $m_V$ is plotted against $m_\pi^2$ for
all three sets. Black circles represent the ground state, $\rho$, and red
squares represent the first excitation, $\rho'$. The experimental values are
depicted as magenta crosses with decay width indicated. R.h.s.: The ratio
$a_V/a_T$ is plotted against the smearing width $R$ for all three data sets.
Black circles represent the ground state and red squares the first excitation.
Broken lines are drawn only to guide the eye  (color online).}
\label{fig:masses}
\end{center}
\end{figure}

We include in our cross-correlation matrix the four interpolators
\bey
&O_1=\overline u_n \,\gamma^i \,d_n\ ,\qquad
&O_2=\overline u_w \,\gamma^i\, d_w\ ,\\
&O_3=\overline u_n \,\gamma^t \,\gamma^i \, d_n\ ,\qquad
&O_4=\overline u_w\, \gamma^t \,\gamma^i \, d_w\ .
\eey
Here $\gamma^i$ denotes one of the spatial Dirac matrices and $\gamma^t$ the
$\gamma$-matrix in (Euclidean) time direction. The subscripts $n$ and $w$ (for
narrow and wide) denote the two smearing widths, $R\approx 0.34$ fm  and $0.67$
fm, respectively. With these interpolators we are able to extract both the
ground state mass and the mass of the first excited state of the $\rho$-meson,
see the l.h.s.\ of Fig.\ \ref{fig:masses}.

On the r.h.s.\ of Fig.\ \ref{fig:masses} we show the $R$-dependence of the ratio
from Eq.\ \eqref{eq:ratio} for the case $a_V/a_T$ both for the ground state
$\rho$-meson and its first excited state. This ratio of the vector to the
pseudotensor coupling to the states shows us their decomposition in terms of the
$(0,1)\oplus (1,0)$ and $(1/2,1/2)_b$ representations. For the ground state at
the largest value of $R\approx 0.67$ fm this ratio is approximately $1.2$,
i.e., we see a strong mixture of the two representations in the wave function
of the ground state $\rho$-meson. Inverting the unitarity transformation from
Eq.\ \eqref{eq:trafo} results in the fact that the vector meson is predominantly
a ${}^3S_1$ state with a tiny admixture of a ${}^3D_1$ wave, $0.997\vert
{}^3S_1\rangle - 0.073 \vert {}^3D_1\rangle$.  This result indicates that the
$\rho(770)$ at the scale fixed by the meson size is approximately a ${}^3S_1$
state -- in agreement with the quark model language.

However, the situation changes for the first excited state, $\rho' =
\rho(1450)$. In this case a strong dependence of the ratio on the infrared scale
is observed. Extrapolating the results to the scale of the $\rho'$ size, $R \sim
0.8 - 1$ fm, one expects a significant contribution  from the $(1/2,1/2)_b$
representation and a contribution of the other representation is suppressed.
This indicates a smooth onset of effective chiral restoration.

The interpretation is as follows. From the conformal symmetry of QCD one expects
that in the deep ultraviolet the pseudotensor interpolator decouples from  the 
$\rho$-mesons. This can also be seen from the non-vanishing anomalous dimension
of the pseudotensor operator, implying its decoupling in the ultraviolet limit.
Thus, the ratio $a_V/a_T$ must increase for small $R$. At large $R$ the ratio
determines a degree of  chiral symmetry breaking in the infrared region, where
mass is generated. 

In the $\rho(770)$ meson chiral symmetry is strongly broken since this state is a
strong mixture of $(0,1)\oplus (1,0)$ and $(1/2,1/2)_b$ with approximately equal
weights.  Consequently, its ``would-be chiral partners'' have a much larger mass:
$a_1(1260)$ and $h_1(1170)$. To these low lying states we cannot assign any
chiral representation. For the $\rho(1450)$ the contribution from $(1/2,1/2)_b$
is much bigger than the contribution of the other representation. One then
predicts that in the same energy region there must exist an $h_1$ (and not an
$a_1$) meson as an approximate chiral partner of $\rho(1450)$. And in fact there
is a state $h_1(1380)$ and no $a_1$ state in the same energy region. The second
excited $\rho$-meson, the $\rho(1700)$, should then be dominated by the
representation $(0,1)\oplus (1,0)$. This assumption is favored by the existence
of the $a_1(1640)$ state. There is no room for this $a_1(1640)$ meson within the
the quark model \cite{PDG,IS}.

Although we do not have the precise value of the ratio $a_V/a_T$ for
$\rho(1450)$ at large $R \sim 0.8 - 1$fm,  it is indicative that this value is
small. Then we are able to give a qualitative estimate for the angular momentum
content of the $\rho(1450)$ in the infrared. Assuming a vanishing ratio  the
state would have the following partial wave content,
\be
\sqrt{\frac{1}{3}}\, \vert {}^3S_1\rangle - \sqrt{\frac{2}{3}}\, \vert
{}^3D_1\rangle\ .
\ee
This shows a leading contribution of the ${}^3D_1$ wave. Possible small
deviations of the ratio from zero do not change this qualitative conclusion.
This result is inconsistent with $\rho'$ to be a radial excitation of the ground
state $\rho$-meson, i.e., an ${}^3S_1$ state, as predicted by the quark model
\cite{PDG,IS}.

\acknowledgments

\noindent
We thank G.\ Engel and C.\ Gattringer for discussions. L.Ya.G.\ and M.L.\
acknowledge support of the ``Fonds zur F{\"o}rderung der Wissenschaftlichen
Forschung'' (P21970-N16, DK W1203-N08) and DFG project SFB/TR-55, 
respectively. The calculations have been performed on the SGI Altix 4700 of the
Leibniz-Rechenzentrum Munich and on local clusters at the ZID at the University
of Graz.

\end{document}